\documentstyle[aps,twocolumn,epsfig]{revtex}
\input epsf
\newcommand{\be}{\begin{eqnarray}}
\newcommand{\ee}{\end{eqnarray}}
\begin{document}

\twocolumn[\hsize\textwidth\columnwidth\hsize\csname @twocolumnfalse\endcsname
\title{High energy collisions of strongly deformed nuclei:\\
 An old idea with a new twist }
 \author{ E.V.~Shuryak}
\address{State University of New York, 
       Stony Brook, NY 11794, USA}
\maketitle
\begin{abstract}
UU collisions can provide about 30\% 
larger densities compared to central PbPb ones. New aspect is
generation of rather deformed initial states.
We show that those can be effectively used to resolve a number
of outstanding issues, from corrections to hard processes, elliptic
flow
(the QGP push issue), and the mechanism of $J/\psi$ suppression. 
UU collisions 
are studied by a simple Monte-Carlo model, and
it is shown how selecting two control parameters -
the number of participant nuclei and deformation - one can select
particular
geometry of the collision.  
\end{abstract}
\vspace{0.1in}
]
\begin{narrowtext}   
\newpage 

\section{Introduction}
An ``old idea'' mentioned in the title is to
select the head-on ``long-long'' collisions, by simply triggering on maximal
number of participants $N_p$. Although it is
kind of a folklore of the field, the only written
material on it I found is a memo \cite{PBM}
  written by P.Braun-Munzinger. His estimates show that, due to
 larger A and $deformation$, the gain in energy density for UU over
 AuAu\footnote{Below we use PbPb instead.}   
can reach the factor 1.8. Although our study had found smaller
numbers, new applications are proposed. They are mostly related with a
different geometry of the collisions, the parallel one, which
incorporates about the same energy density as available in central
PbPb
collisions, with significant deformation. 

The general attitude of this study was to look how UU collisions can
help
to understand the existing open issues of the SPS heavy ion program.
It certainly is of interest to RHIC program, although it is
probably premature
to discuss it now.
The emphasis is made here on event selection, and other interesting
options (like using targets with crystals with naturally aligned U)
are not studied.

  Let me start with outlining simple argument for head-on collisions.
 Representing U as a homogeneous ellipsoid with one long ($R_l$) and two
short
($R_s$) semi-axis,  one can related their ratio to deformation
parameter
$\delta$ used in nuclear physics (see e.g. \cite{BM})
\be
{R_l \over R_s}=({1+4\delta/3 \over 1-2\delta/3})^{1/2}
\ee
For $\delta_U\approx .27$ this ratio is 1.29, the basic deformation
ratio to be used below.

It is convenient to think first in terms of ``wounded'' or ``participant''
 nucleons first, a purely geometric concept, and only then
consider real multiplicity (entropy) production. (We will follow such
logic throughout the paper.) Let us thus start with comparing 
  the density of
participants per transverse area $n_p=N_p/(\pi R_s^2)$, for
``long-long'' collisions 
of the deformed nuclei vs the spherical one with {\it the same} A and
 $R=(R_s^2 R_l)^{1/3}$. The effect only comes from reduction of the
 area, so
\be
{n_p^{deformed} \over n_p^{spherical}}={A^{deformed} \over A^{spherical}}({R \over R_s})^2
\ee
For A=238 we will use $R_l=8.4, R_s=6.5, R=7.0 fm$, and so deformation
alone increase the $n_p$ by 1.16. For U and Pb ($R_{Pb}$=6.78 fm in
such model) one gets the participant density gain 1.24.

  Transferring this into initial entropy density, one should recall
  that
for  (spherical) AA collisions
\be
{dN \over dy}(y=0) \sim A^{1+\alpha}
\ee
with $\alpha\approx .12$. So, assuming as usual that final
multiplicity is proportional to the initial entropy density,
we see that there is a correction to the simple idea that each
participant
nuclei gives the same (energy dependent) contribution to the spectrum.
This non-zero $\alpha$ incorporates both (i) additional increase in
multiplicity,
and (ii) extra stopping (shift toward mid-rapidity):
we are only interested in their combination  dN/dy(y=0).
Furthermore, it is natural to think that transverse dimensions of the
system enter trivially here, and so these extra effects due to increased
density.  With this additional factor $\sim n_p^{3\alpha}$ we
obtain the UU/PbPb {\it total initial density gain}\footnote{At fixed
  time after the crossing, a la Bjorken. If more
  stopping means earlier equilibration, the available density is larger.}
  $1.24^{1+3\alpha}=1.34$. 

The main questions addressed in this paper are two-fold.
One is  to make some realistic
estimates of the effect, not just for a particular configuration but for
ensemble of events selected by some {\it experimentally accessible}
criteria. The second is 
 to outline possible applications of high energy collisions of the
 deformed nuclei. 

\section{UU collisions versus PbPb: the simulation   }
  Simple Monte-Carlo program was written, which initializes nucleons
  inside
nuclei and follow their paths through another one. Since we are not
really interested in peripheral collisions, we did not included
diffuse boundary of nuclei and used the ellipsoids described above.
We also ignored probabilistic nature of the interaction, considering
transverse distance between nucleons $R<(\sigma_{in}/\pi)^{1/2}$ to be
sufficient reason to make both of them participants.  So, the only
source of fluctuations\footnote{And it is by no means assumed to be
 accurate account
  for fluctuations.} are random positions of the nucleons inside the
nucleus.

 Spherical nuclei have only one parameter - impact parameter b -
which in such classical treatment determines the mean number of
participants. Deformed nuclei have in general 5 such parameters:
b and 4 spherical angles $\theta_i,\phi_i,i=1,2$ indicating
the orientation of their longer axes at the collision moment.
The main objective of the calculation is to see how well
one can actually fixed those, by using experimentally available
information.

  Few words about our definitions. After all participant nuclei are
  identified, we calculated the tensor
\be T_{ij} =<x_i x_j> \ee
in transverse plane, diagonalize it and find its eigenvalues
 $R^2_+,R^2_-$. The density of participants we use below is defined as
$n_p=N_{part}/(\pi R_+R_-)$ and deformation as $R_+/R_-$.  
\begin{figure}[t]
\epsfxsize=3.in
\centerline{\epsffile{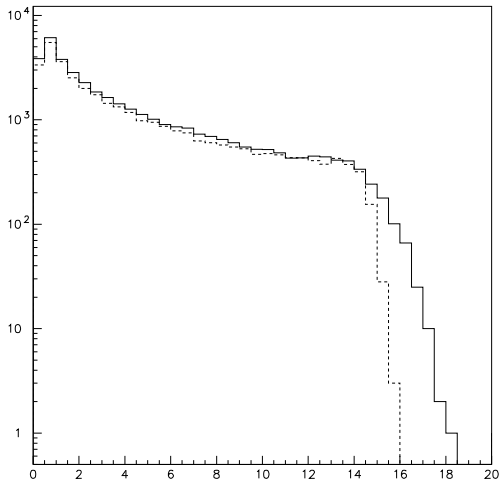}}
\vskip -0.1in
\caption[]{
 \label{fig_density} Density distribution with selection $N_p>0.9 (2A)$.
 Solid (dashed)
are for UU and PbPb collisions }
\end{figure}

If no selection is made, the density of the participants $n_p$ 
 shown in \ref{fig_density} is very
similar, except for the tail region. A gain in $n_p$ 
of the order of 16\% 
seem indeed possible, with reasonable loss of statistics.
However, if one introduces ``centrality'' cuts, the situation
is different. We would define a fair
 ``centrality cut'' by restricting $N_p>0.9 (2A)$, with corresponding
 A for for both cases.  Triggering on large $N_p$ (or forward energy)
one effectively eliminates
 spectators  (and 
many complicated geometries).
\begin{figure}[ht]
\epsfxsize=3.in
\centerline{\epsffile{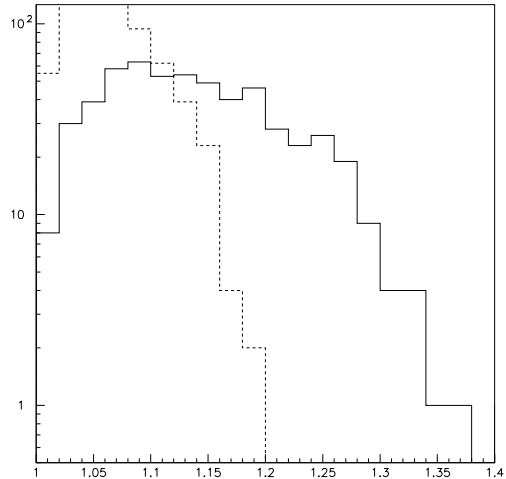}}
\vskip -0.05in
\caption[]{
 \label{fig_ratio}
  Distribution of $R_+/R_-$ (long-to-short semi-axis in transverse plane),
 with selection $N_p>0.9 (2A)$.
 Solid (dashed)
are for UU and PbPb collisions}
\end{figure}
The first striking feature one finds after such cut
is that distribution of the deformations of the initial 2d ellipsoid $R_+/R_-$
 is very
different:
see figure Fig.\ref{fig_ratio}. The ratios as large as 1.35 are
accessible, while in PbPb collisions all the ``central'' collisions are
very spherical. The maximal deformation, not surprisingly, is of the
order of deformation of U\footnote{Note that it
 is comparable to what is obtained for medium b for spherical
nuclei, but now reached for  larger energy density and larger system.
}
Those correspond to collisions with two long directions 
parallel to each other and orthogonal to the beam.
It is about the same as is obtained for medium b for spherical
nuclei, but at larger energy density and larger system (see below).

The joint distributions in participant density - deformation plane 
for (most central) UU and PbPb collisions are compared in
Fig.\ref{fig_denrat}(a,b).
 The main message  one can get from it is that
strong correlation between the deformation and density,
 existing for spherical nuclei,
is to some extent relaxed for UU.

  How one can measure the 2d deformation of the initial conditions?
The measured  elliptic deformation of spectra of secondary particles,
pions or nucleons, $v_2$, is proportional
to this initial deformation (with EOS depending coefficient) 
 and should have similar distribution.
divide measured
 distribution over $v_2$ into more and less deformed.

 Suppose now that one uses both control parameters, $N_{part}$ $and$  $v_2$.
 Is it really possible to
 select the {\it particular geometries} of the
collisions
we want? In  Fig.\ref{fig_theta}(a) one can see that it is to a
significant effect correct: the $less-deformed$ sample is rich in the region
$cos\theta\approx 1$, or in ``head-on'' collisions, while the
$more-deformed$ 
collisions
have none of them, and concentrate at small $cos\theta$. In
Fig.\ref{fig_theta}(b)
one can see that the same selection of events correspond
to the {\it difference} of the azimuthal angles $\phi_1-\phi_2$
to be peaked around 0, or have a flat distribution, respectively.

\begin{figure}[t]
\epsfxsize=3.in
\centerline{\epsffile{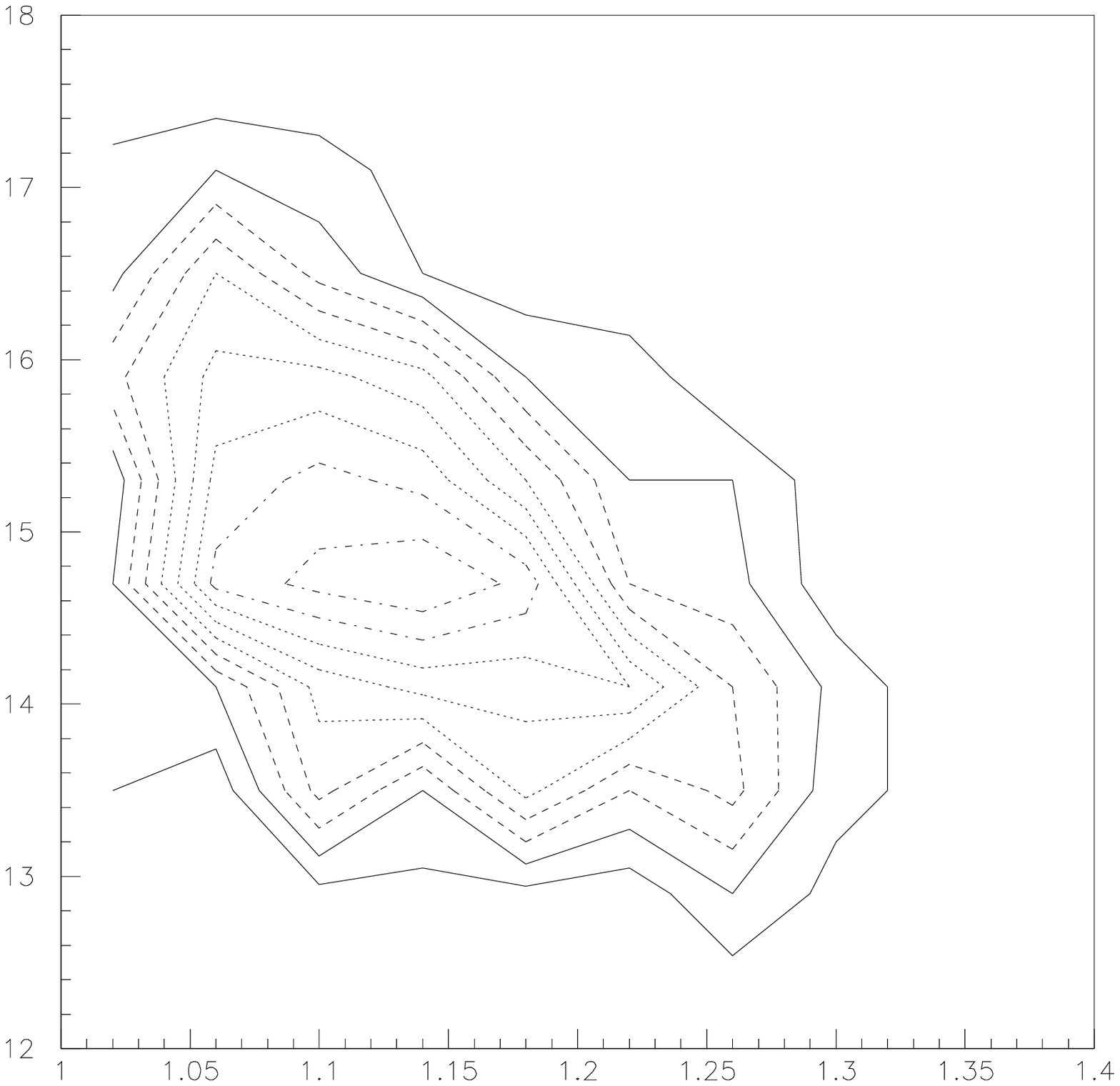}}
\epsfxsize=3.in
\centerline{\epsffile{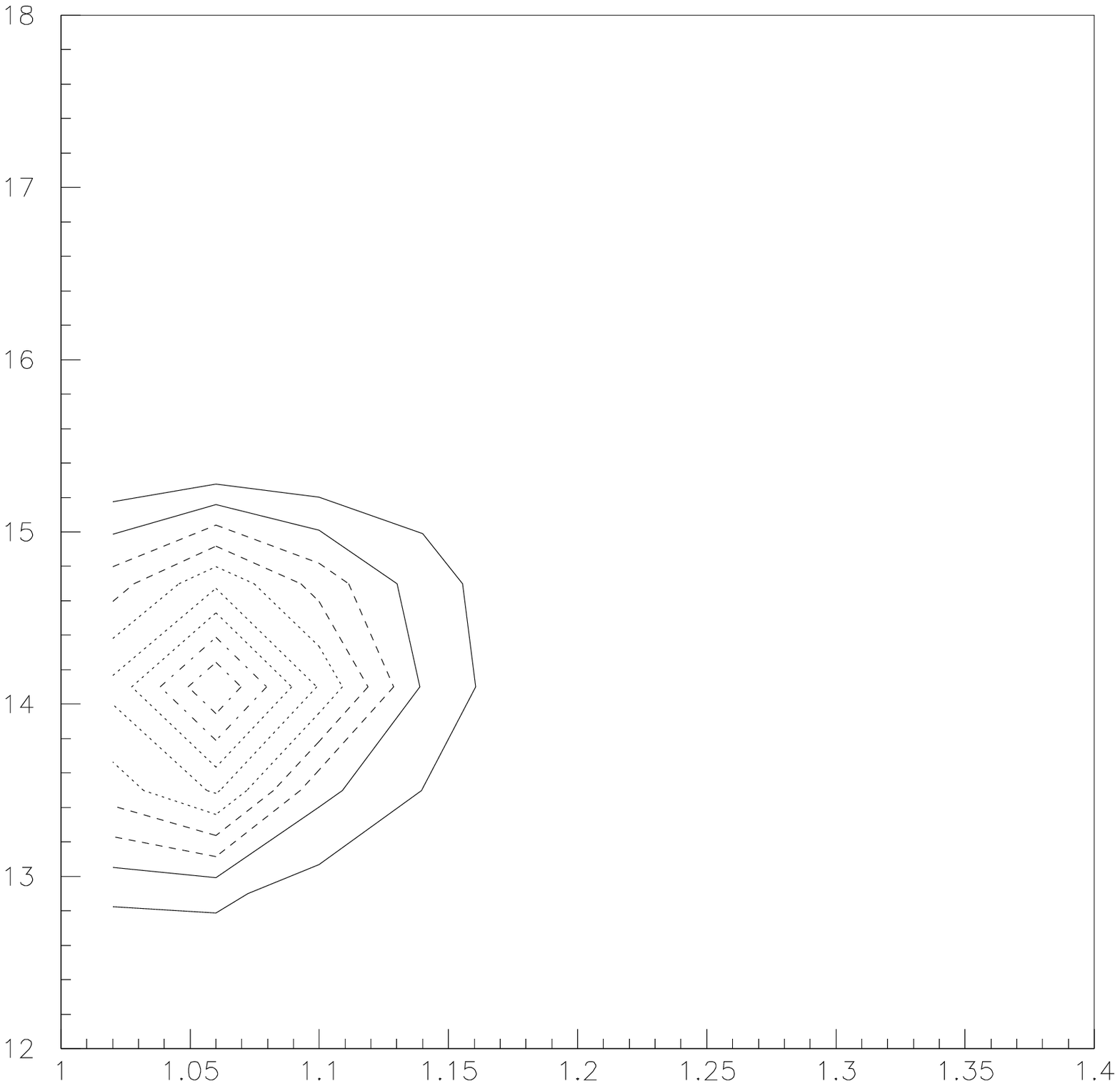}}
\vskip -0.05in
\caption[]{
 \label{fig_denrat}
Distribution over
participant density $n_p$ $[fm^{-2}]$ vs deformation $R_+/R_-$, for (a)
 UU ($N_{part}>428$) and (b)) PbPb  ($N_{part}>374$) collisions, respectively. }
\end{figure}

\begin{figure}[t]
\epsfxsize=3.in
\centerline{\epsffile{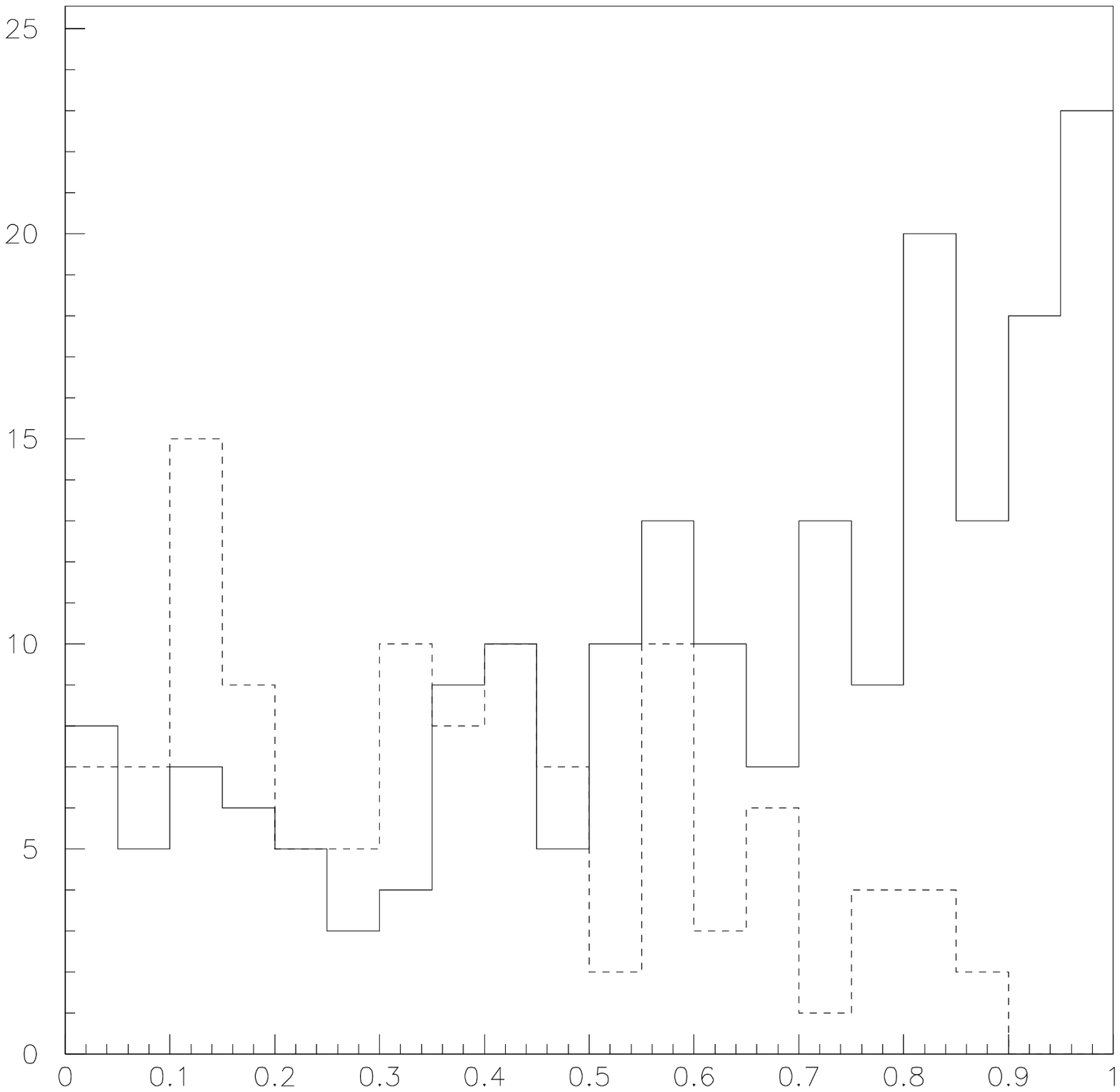}}
\epsfxsize=3.in
\centerline{\epsffile{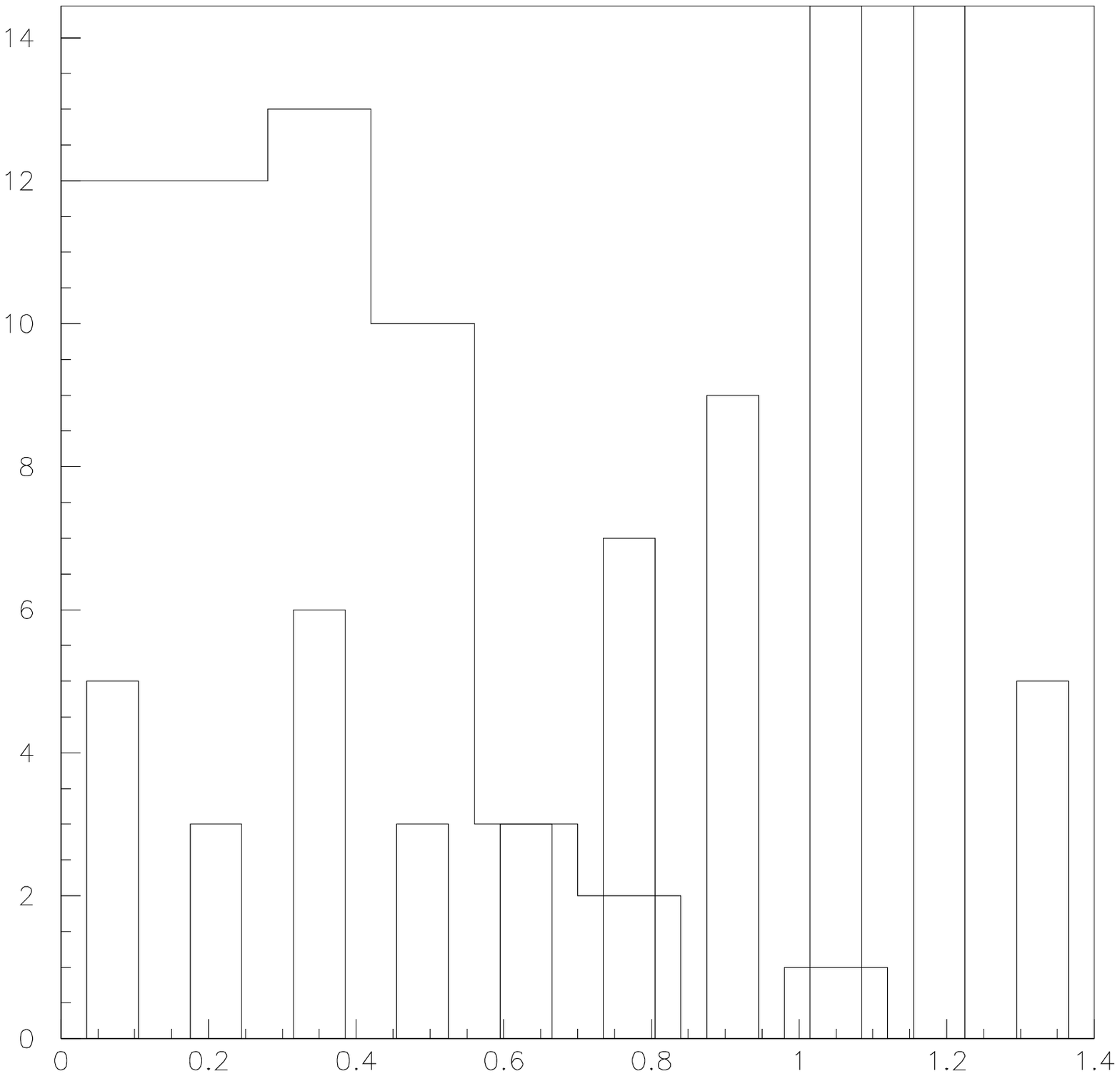}}
\vskip -0.05in
\caption[]{
 \label{fig_theta}
Distributions of original angles for UU collisions, $N_p>0.9 (2A)$.
  (a) Distribution in  $cos(\theta)$,  the angle between the long axis of U
and the beam.  The solid (dotted) histograms are for less deformed
$R_+/R_-<1.1$ ( more deformed $R_+/R_->1.2$) initial states  .
 (b)Distribution in difference between polar angles $|\phi_1-\phi_1|$,
  solid (barred) histograms are for less and more deformed collisions, same
selection.}
\end{figure}

Finally note also that
the figures presented above 
have shown only the density of participants $n_p$: let
us now recall the correction factor $\sim n_p^{3\alpha}$  we discussed
in the Introduction for the initial energy density. It can be seen as
 additional non-linear deformation of the axis, when going from $n_p$
 to   dN/dy. The contrast between  UU and PbPb in  dN/dy is larger
 by another 8\%.  
Note also, that we have only discussed local quantities, like participant
density $n_p$. For many applications the absolute size of the system
is of similar and sometimes even larger importance.

\section{Possible applications}
\subsection{Hard  versus  soft processes}
The so called $hard$ processes include Drell-Yan dilepton production, two-jet events,
heavy flavor production etc. Those are described in the first order by the
parton model, and so
are the simplest to treat for any geometry. For example, for head-on
collisions
discussed in the introduction, the output should be simply
\be \sigma_{hard} \sim R_z^2 R_t^2\ee

 Significant interest is related with various QCD $corrections$
 to the parton model: some of them are
related to {\it initial state} re-scattering, and some to the {\it final state}
ones. Examples of the former are ``shadowing'' of nuclear structure
functions and increase in parton $p_t$, the latter processes
lead to the so called ``jet quenching'' or energy losses.
The debates about their relative contributions
 continues: one, at one talk as a semi-joke I proposed to consider
 rectangular
nuclei with different $R_z,R_t$ to separate them . Now we propose a particular
realization of this idea.

How large level arm do we have, with UU collisions?
For head-on collisions (discussed above) $R_z=R_l.R_t=R_s$, while
for ``parallel collisions'' with maximal deformation it is the other
way around, $R_z=R_s$ and $one$ of the $R_t=R_l$. So the ratio $R_z/R_t$
varies between about 1.3 and 1/1.3, or a span of 1.7.
This may be enough to disentangle different mechanisms.

\subsection{Ellipticity and EOS}
In high energy collisions
the shape of the ``initial almond'' for non-central
collisions leads to enhanced ``in-plane'' flow, in direction of
the impact parameter \cite{Ollitrault}. 
It is very
 important because (as  pointed out 
in \cite{Sorge}(a)) it is developed $earlier$
than the radial one, and thus it may shed  light on Equation of State
(EOS)
at early time. The particular issue is
 whether we do or do not
have QGP at such time, at SPS or RHIC.
Recent review of ellipticity one can find in \cite{QM99_elliptic}.

 Ellipticity is now measured by
the asymmetry of the particle $number$, or 
$v_i$ harmonics defined as
\be
	{dN \over d\phi} = {v_0	\over 2\pi } + {v_2 \over \pi } \cos( 2\phi ) + 
				   {v_4 \over \pi }  \cos( 4\phi ) +
                                   \cdots \ee
rather than asymmetry of the momentum distribution, which was used
more at low energies. 
Furthermore, $v_i$ are often additionally
normalized to the spatial asymmetry of the initial state (the
``almond'')
at the same b, $\alpha_2=(R_+^2-R_-^2)/(R_+^2+R_-^2)$
in order to cancel out this kinematic factor and to see
the response to asymmetry. 

 There are two ways to look at elliptic flow: by using
  collision  energy dependence, or centrality dependence at fixed
  beam.
The first method is more difficult, but it deals with fixed geometry
and only slowly changing size of the system and relevant densities.
For recent discussion of it see ref.3.
 The $centrality$ dependence is easier to measure, but
 interpretation  is more complicated, because increasing b we make
``almond'' more elliptic, but also much smaller and thinner: eventually
finite size corrections reduce pressure build-up.
This is of crucial importance for the ``QGP push'' issue at SPS,
 since it should
be
present only for the largest densities available, and at small b
the $v_2$ is very small and difficult to measure.

 Preliminary 
NA49 data presented at QM99\cite{NA49} may 
indeed indicate ``the plasma push'', as  enhancement of $v_2$ at small
b. However 
 much more work is clearly needed to understand this complex interplay of
EOS and finite size effects.

 U collisions discussed below in principle
provide the means to decouple finite size and deformation issues.
In particularly, the deformation of U (about 1.3) is actually enough
to generate well measurable $v_2$ of the order of several percent,
without
significant loss in density. 

\subsection{$J/\psi$ suppression}

  One of the most intriguing observables, the $J/\psi$ suppression,
is unique in its significant centrality and A-dependence. 
Here is not a place to discuss it in any details, but let 
me make few remarks about recent developments.

New NA50 data reported at QM99 have clarified the situation
for the most central collisions: using now very thin target,
it was found that 1996 data suffered from multiple interactions.
In fact there is a significantly stronger suppression at small b.
Furthermore, it seem like the two component picture,
 with separate $\chi$ and $\psi$
thresholds,
really emerges. In view of this,
it is desirable to increasing the
 density and/or the famous variable L:
only deformed U provides an opportunity here at SPS.

  In order to discriminate
 experimentally different ideas on the nature of {$J/\psi$
   suppression, we should be able to
tell whether suppression happened quickly or need a longer time.
  The old idea is to study suppression dependence on $p_t$.
Unfortunately changing $p_t$ we  also change the kinematics:
e.g. destruction by
   gluons or hadrons go better if  $p_t$ grows.

  Maybe better idea \cite{HeiMat} is to 
  use azimuthal dependence of the suppression.
 Instantaneous suppression should show $no$ asymmetry,
but if it takes few fm/c
  the anisotropy should show up. However, as for the
elliptic flow, the problem is the
initial ``almond'' at b$<$ 8 fm is not very anisotropic,
and for larger b there is no anomalous suppression.

Here too the deformed U can help, providing (with trigger conditions
discussed above) a variety of geometries, including the ``parallel collisions''.

\section{Acknowledgements} 
I thank W.Henning, whose question forced me to think about this
subject, and P.Braun-Munzinger and T.Ludlam for helping me retrieve the ref.1.
This work is supported by US DOE, by the grant No. DE-FG02-88ER40388.

}
\end{narrowtext}

\end{document}